\documentclass[aps, prl, twocolumn, showpacs, superscriptaddress]{revtex4-1}

\usepackage[english]{babel}
\usepackage{graphicx}
\usepackage{wrapfig}
\usepackage{floatrow}
\usepackage{sistyle}
\usepackage{prettyref}
\usepackage{graphicx}
\usepackage{dcolumn}
\usepackage{bm}
\usepackage{amsmath}
\usepackage{times}
\usepackage{color}

\usepackage{natbib}

\begin{document}
\title{Strong paramagnon scattering in single atom Pd contacts}
\date{\today}

\author{V. Schendel}
\email[Corresponding author; electronic address:\ ]{verena.schendel@kit.edu}
\affiliation{Max-Planck-Institut f\"ur Festk\"orperforschung, Heisenbergstra\ss e 1, 70569 Stuttgart, Germany}

\author{C. Barreteau}
\affiliation{Service de Physique de L'Etat Condens\'e (SPEC), CEA, CNRS, Universit\'e Paris-Saclay, CEA Saclay 91191 Gif-sur-Yvette Cedey, France }
\affiliation{ Dept. of Micro and Nanotechnology, Tech. Univ. of Denmark, {\O}rsteds Plads build. 345B, DK-2800 Kgs. Lyngby, Denmark}

\author{M. Brandbyge}
\affiliation{ Dept. of Micro and Nanotechnology, Tech. Univ. of Denmark, {\O}rsteds Plads build. 345B, DK-2800 Kgs. Lyngby, Denmark}

\author{B. Borca}
\affiliation{Max-Planck-Institut f\"ur Festk\"orperforschung, Heisenbergstra\ss e 1, 70569 Stuttgart, Germany}

\author{I. Pentegov}
\affiliation{Max-Planck-Institut f\"ur Festk\"orperforschung, Heisenbergstra\ss e 1, 70569 Stuttgart, Germany}
\author{U. Schlickum}
\affiliation{Max-Planck-Institut f\"ur Festk\"orperforschung, Heisenbergstra\ss e 1, 70569 Stuttgart, Germany}
\author{M. Ternes}
\affiliation{Max-Planck-Institut f\"ur Festk\"orperforschung, Heisenbergstra\ss e 1, 70569 Stuttgart, Germany}
\author{P. Wahl}
\affiliation{SUPA, School of Physics and Astronomy, University of St. Andrews, Scotland, United Kingdom}
\affiliation{Max-Planck-Institut f\"ur Festk\"orperforschung, Heisenbergstra\ss e 1, 70569 Stuttgart, Germany}
\author{K. Kern}
\affiliation{Max-Planck-Institut f\"ur Festk\"orperforschung, Heisenbergstra\ss e 1, 70569 Stuttgart, Germany}
\affiliation{Institut de Physique, Ecole Polytechnique F\'ed\'erale de Lausanne, 1015 Lausanne, Switzerland}

\begin{abstract}
Among all transition metals, Palladium (Pd) has the highest density of states at
the Fermi energy yet does not fulfill the Stoner criterion for ferromagnetism. However, its close vicinity to magnetism renders it a nearly
ferromagnetic metal, which hosts paramagnons, strongly damped spin fluctuations. In this letter we compare the total and the differential conductance of mono-atomic Pd and Cobalt (Co) contacts between Pd electrodes. Transport measurements reveal a conductance for Co of 1\,$G_{0}$, while for Pd we obtain 2\,$G_{0}$.
The differential conductance of mono-atomic Pd contacts shows a drop with increasing bias, which gives rise to a peculiar
$\Lambda$-shaped spectrum. Supported by theoretical calculations we correlate this finding with the life time of hot quasi-particles in Pd which
is strongly influenced by paramagnon scattering. In contrast to this, Co adatoms locally induce magnetic order and transport through single cobalt atoms remains unaffected by paramagnon scattering, consistent with theory.


\end{abstract}

\maketitle

Spin fluctuations are believed to provide the pairing glue in unconventional
superconductors \cite{FayPRB, Moriya, mathur_magnetically_1998, scalapino_common_2012}. Therefore, the interaction of magnetic
fluctuations with electronic degrees of freedom is critical for a
full understanding of unconventional superconductivity. A material which in its
elemental form exhibits strong magnetic fluctuations yet does not even become a
conventional superconductor is palladium \cite{Stritzker}. This raises important questions as to
how spin fluctuations interact with the conduction band electrons.

Spin fluctuations also play an important role in some of the macroscopic properties of elemental metals:
Both, palladium (Pd) and platinum (Pt) are not ferromagnetic despite an only partially filled $d$-shell, but belong to a class of materials coined \textit{nearly} ferromagnetic metals \cite{Lonzarich1985}. Pd possesses the
highest density of states (DOS) at the Fermi energy ($E_{\mathrm F}$) among all transition metals
and the Stoner criterion is almost fulfilled \cite{VanLeeuwen1992} bringing it right to the
edge to ferromagnetism. 
In these nearly ferromagnetic metals, strongly damped spin fluctuations, known as
paramagnons, have a great impact on macroscopic quantities such as the heat
capacity and magnetic susceptibility \cite{Mueller1970}. Paramagnons in Palladium have been commonly observed by means of scattering techniques such as neutron scattering \cite{Doubble2010} and angle-resolved photoemsission \cite{Hayashi2013} within an energy range of about 50-150\,meV\,-\,detection on the atomic level, however, remained illusive.



Paramagnons can be described as magnetic fluctuations of a paramagnetic phase
\cite{sushkov2014}. In contrast to magnons, which are the fluctuations of a
magnetically ordered system, paramagnons are collective overdamped modes with
only short correlation lengths that appear close to a magnetic instability
\cite{Lonzarich1985, Moriya1978}.
Recent experimental \cite{Teng2008, Xiao2009,
VanLeeuwen1992, Rodrigues2003, Ienaga2015} and theoretical \cite{Lee1998,
Delin2003, Smelova2008, Kudasov2007, Sun2010, Gava2010} studies have explored
the possibility of Pd becoming ferromagnetic in nanostructures.
In this work we study transport through single Pd and Co adatoms on a
Pd(111) surface by scanning tunneling microscopy (STM).
Differential conductance ($dI$/$dV$) spectra were taken at different tip-sample
distances $z$ from the tunneling to the contact regime. In Pd contacts, we find
that the spectral features show a significant decrease of the conductance with
increasing bias, independent of the polarity. Theoretical calculations show
that this feature can be correlated to the extremely short lifetime of hot
quasi-particles in Pd -- an effect that we attribute to paramagnon excitations. For
Co contacts, the differential conductance is comparatively featureless.


Experiments have been performed on a Pd(111) single crystal with a home-built
UHV-STM operating at 6\,K. The sample was cleaned in vacuum with a base pressure of
$3 \cdot 10^{-10}$\,mbar by numerous cycles of Ar$^{+}$ sputtering and
subsequent annealing. The most frequent bulk contaminants in Pd are
sulphur (S) and carbon (C). Upon annealing to temperatures of 1000\,K for
extended periods, S and C impurities migrate to the surface.
S is removed by sputtering, while C is removed by exposing the crystal to an oxygen
atmosphere ($p_{\mathrm O2}$= 3.0\,$\cdot$10$^{-7}$\,mbar) for 20\,min while
heating the sample to temperatures in the range of 650-850\,K. The final
preparation cycle was carried out in the absence of oxygen and the sample was
annealed to 900\,K. The apex of the STM tip was covered with Pd by
gentle indentation into the surface.
Single Pd atoms were released from the tip by approaching the tip towards the bare
surface until contact is formed [Fig.~1(a)]. Co atoms were evaporated from a wire with 99.99\% purity in-situ onto the sample being held at 6\,K [Fig.~1(b)]. Both species can be easily distinguished from each other by their apparent height [Fig.~1(c),(d),(e)]


\begin{figure}
\onecolumngrid
	\centering
	\includegraphics[width=\columnwidth]{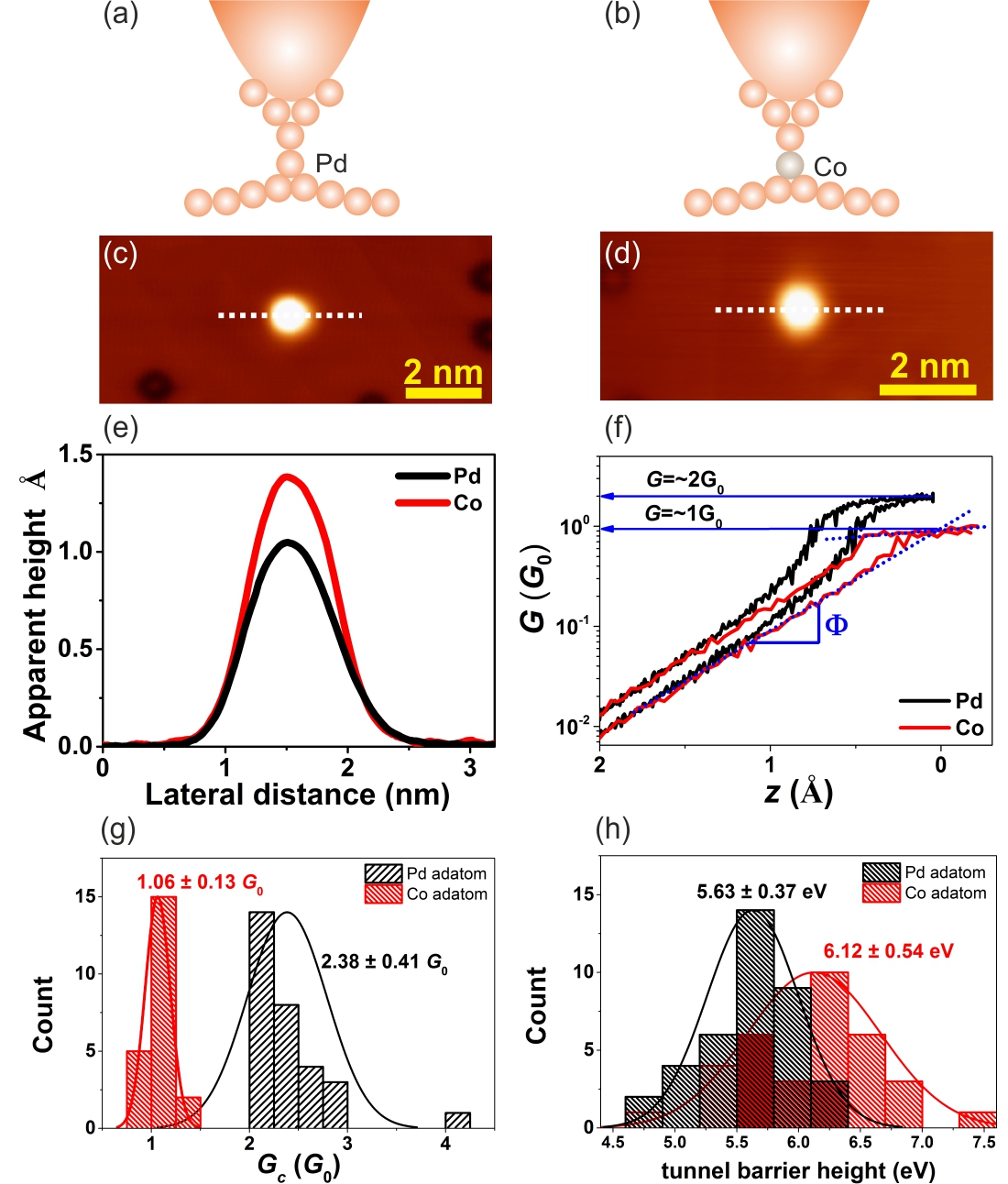}
	\caption{(a),(b) Sketch of mono-atomic Pd (a) and Co (b) contacts. (c),(d)
Constant current images of a Pd (c) and a Co (d) adatom acquired
on Pd(111) at 6~K ($V=0.1\mathrm V$, $I=100\mathrm{pA}$). (e) Apparent height
profile of a single Pd and Co atom.  f, Conductance-displacement ($G(z)$)
curves taken on Pd (solid black line) and Co (solid red line). Mono-atomic Pd
contacts show a characteristic conductance of about $2G_{0}$ while Co contacts
exhibit ones of about $1G_{0}$. (g),(h) Conductance and tunnel barrier height measurements on Pd and Co adatoms deposited onto Pd(111). To extract the mean values, the histograms were fitted with Gaussian functions. }
	\label{fig1}
\end{figure}


Distance dependent conductance, $G(z)$, measurements have been carried out
for contacts between the tip and individual Pd and Co adatoms. From the $G(z)$
curves two regimes are discernible; a tunneling and a contact regime [Fig.~\ref{fig1}(f)]. In the tunneling regime, a decrease in $z$ is associated with an
exponential increase of the conductance, i.\,e.~$G(z)=G_0\exp(-2\kappa z)$ (with $G_0=2e^2/h=77.5 \mu\mathrm S$ as the quantum of conductance).
The slope $\kappa=\sqrt{\frac{m_0}{\hbar^2}\Phi}$ is directly related to the
local tunnel barrier height $\Phi$ \cite{Binnig1982}. Reducing $z$ further
leads to a relaxation of the tip and the surface atoms due to adhesive forces.
When contact between the tip apex atom and the surface is established, a
discontinuous jump in $G(z)$ occurs. This jump arises when the bonding
strength between surface and tip apex atom overcomes that between the atoms
within the tip. As displayed in Fig.~\ref{fig1}(f) for the approach curve on a Co
atom, the conductance of the contact $G_c$ can be obtained by extrapolating the
tunneling regime to the intersection with the contact regime, which is defined
as $z=0$. Positive $z$ values denote tunneling, whereas negative ones
the contact regime.

The measured values of $G_c$ and $\Phi$ for both types of adatoms are summarized in Fig.~\ref{fig1}(g) and
(h).
For the tip-Pd adatom and tip-Co adatom contacts we found
$G_c=2.38\pm0.41\,G_{0}$ and $1.06\pm0.13\,G_{0}$, respectively.
The results for Co adatoms are consistent with previous reports for
Co adatoms on noble metal surfaces \cite{Neel, Vitali}. For Pd contacts, the
reported values exhibit variations depending on preparation conditions.
For break junctions prepared in vacuum at room temperature, a conductance of 0.5$G_0$ was found and interpreted in terms of a single spin-polarized conductance channel \cite{Rodrigues2003}. Other measurements carried out on Pd break junctions at low temperatures and
in vacuum reported conductances in good agreement with our values\cite{Csonka2004, Matsuda2007}.



Differential conductance (d$I$/d$V$) spectra were obtained on top of Pd and Co adatoms starting from the tunneling regime to contact [Fig.~2(a),(b)]. Spectra taken on the Pd adatoms in the tunneling regime give access to the distribution of occupied and unoccupied states near $E_{\mathrm F}$, hence reflecting the LDOS of Pd \cite{Mueller1970}. However, when the contact regime is reached a drastic change of the feature to a distinct $\Lambda$-shaped peak occurs. Further increasing the setpoint $G$ leads to a significant broadening and decrease of the signal strength of the $\Lambda$-like feature. Contrary to these observations the d$I$/d$V$ spectra taken on Co adatoms show a rather flat signal in the tunneling as well as in the contact regime. The small feature on the cobalt atoms around zero bias might be due to emergent Kondo correlations, which are commonly observed for magnetic adatoms on noble metal surfaces \cite{Ternes2009a,Wahl2005}.

\begin{center}
\begin{figure}[h]
	\centering
	\includegraphics[width=\textwidth]{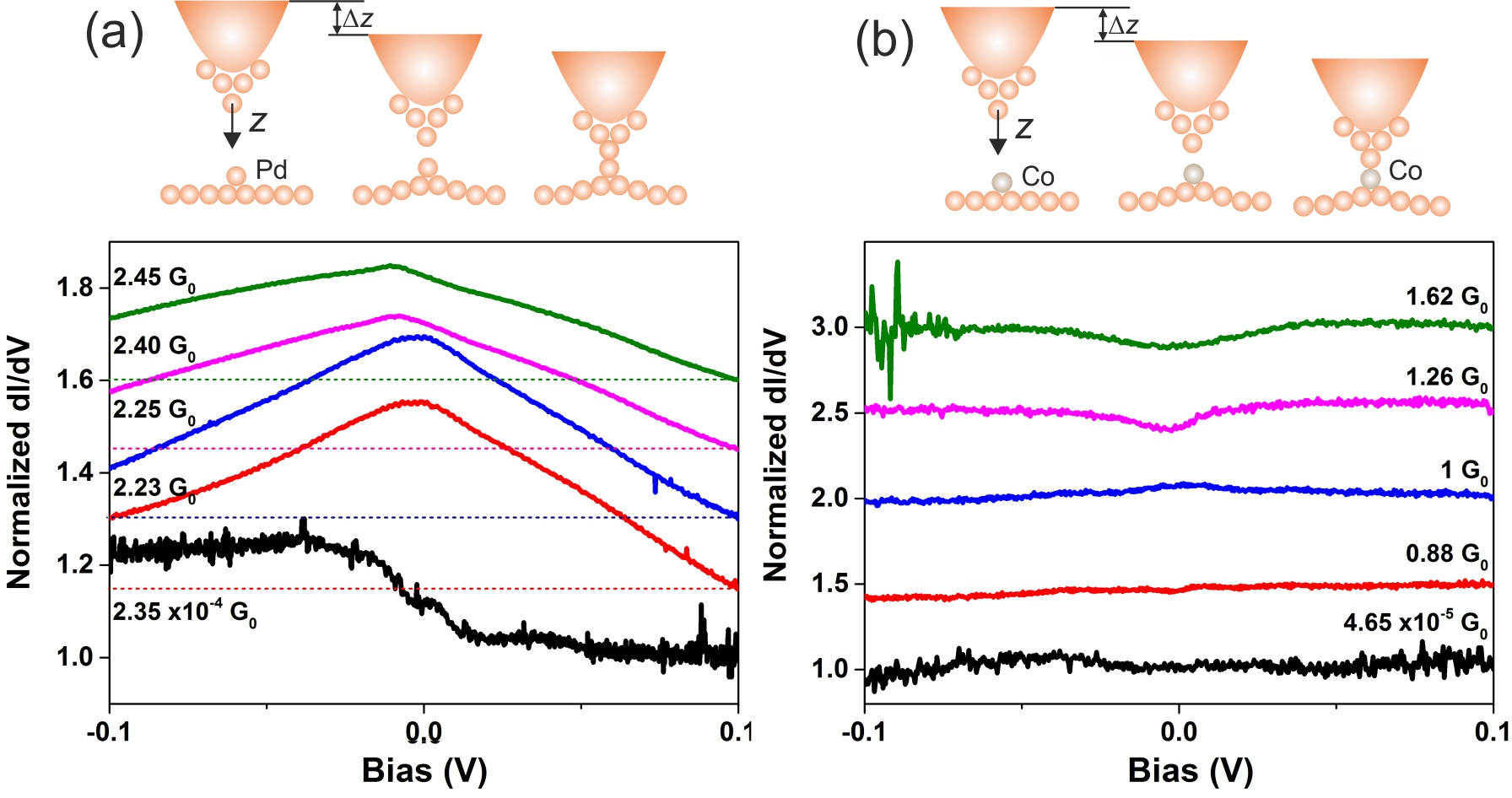}
	\caption{Differential conductance (d$I$/d$V$) acquired on a Pd adatom (a) and Co adatom (b) deposited on Pd(111). Spectra were recorded with a lock-in modulation of 2\,mV and taken through different heights from tunneling to contact as indicated by the setpoint conductance. Curves are normalized at $V=0.1\mathrm{V}$ (horizontal dashed lines) and stacked for clarity. 
	}
	\label{fig3}
\end{figure}
\end{center}

To understand the $\Lambda$ anomaly, we have performed  Density Functional Theory (DFT) and electronic transmission calculations in the Non Equilibirum Green
Function (NEGF) formalism with the Atomistix Toolkit (ATK) code from QuantumWise\cite{quantumwise} and the Transiesta code\cite{Brandbyge2002} to model the transport through single cobalt and palladium adatoms. All our calculations were performed within the local density approximation (LDA) using the Perdew Zunger parametrization \cite{Perdew1981} since the generalized gradient approximation incorrectly predicts a magnetic ground state for Pd \cite{Alexandre2006b}\footnote{Our LDA calculations predict an equilibrium lattice parameter of $3.87$\,{\AA} for face centered (fcc) Pd and an on-set of magnetism for lattice parameters above $3.96$\,{\AA} while GGA calculations predict a magnetic bulk at equilibrium.}.
We used a single zeta polarized (SZP) basis set and separable norm-conserving Troullier-Martins pseudopotentials\cite{Troullier1991}  with partial core corrections.

We have performed DFT calculations for these two systems to establish the differences in the electronic states near $E_\mathrm F$. We found no evidence of spin-polarization for the Pd adatom, while a strong spin polarization develops on the Co adatom with a spin moment of $2.76 \mu_\mathrm B$. In addition, we find non-negligible spin-polarization on neighboring Pd atoms, indicating that the Co atom locally induces magnetic order [Fig.~3(a)].

Fig.~3(b) depicts the projected density of states (PDOS) of the $d$ orbitals for a bulk and surface Pd atom as well as for the adatom (Pd or Co). The $d$ band of Pd is almost filled and $E_\mathrm F$ falls into the tail of the $d$-states\footnote{Note that when considering the effect of a tip in contact with the adatom we have found that a rather pronounced dip develops around $E_\mathrm F$ which will play strongly against the magnetization of the adatom.}.
For Cobalt, the $d$-states exhibit a strong exchange splitting. The states at $E_\mathrm F$ are dominated by $d$-states of minority spin character.

\begin{figure}
\onecolumngrid
	\centering
	\includegraphics[width=7cm]{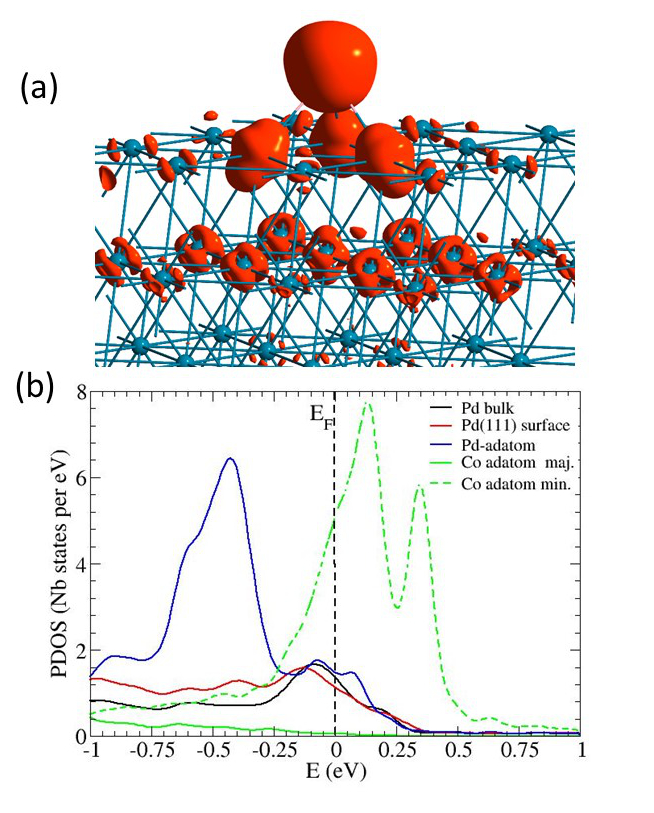} 
	\caption{\label{PDOS}
(a) Real-space distribution of the spin-density isosurface plot of a Co adatom on a Pd(111) surface. The magnetic moment on the cobalt atom is $2.76\mu_{\mathrm B}$, while the one induced on the palladium atoms is $0.26\mu_{\mathrm B}$ on the nearest neighbours and $0.13\mu_\mathrm B$ in the sub-layer.	
(b) Calculated projected density of states (PDOS) on the $d$ orbitals of a bulk (black), $(111)$ surface atom (red), Pd (blue) and Co (green) adatom. In the case of the Cobalt adatom the majority spin (full green) and the minority spin (dashed green) are splitted by a large exchange and the principal PDOS contribution from the majority spin is well below $E_F-1$eV while for palladium there is no magnetism and the two spins are degenerate. }
\end{figure}

A series of electronic transport calculations was carried out where the system is divided into three regions: left and right leads and central region containing the atomic contact. The leads are built from a semi-infinite repetition of 3 atomic layers with an fcc stacking. A $4\times4$  unit cell is used and periodic boundary conditions are applied in the $(111)$ plane. We checked that a $5\times 5$ unit cell did not change our results significantly. The central part is made of 3 layers in contact via a 4 atom pyramid the apex of which is at a distance $d$ from the adatom. Only the adatom and the pyramid have been allowed to relax. A sketch of the system is presented in the inset of Fig.~4(a).

\begin{figure}

	\medskip
	\vspace{-12pt}
	\centering
	\includegraphics[width=7cm]{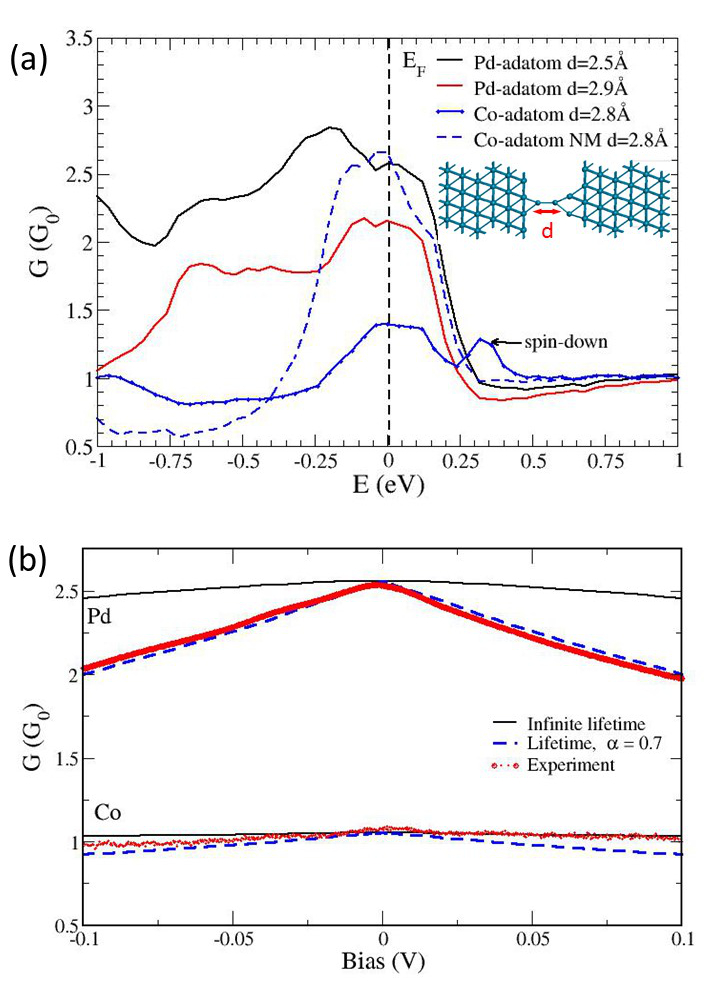}
	\caption{
 (a) Calculated electronic transmission between a tip and a  monatomic Pd (Co) adatom. $d$  is the distance between the adatom and the apex of the tip. In the case of the Co adatom the system is spin polarized. Additionally we have also considered the case of a non-magnetic (NM) Co adatom. The sharp feature above 0.25\,eV (indicated by an arrow) originates from the spin-down contribution. The Fermi level is set as the zero of energy.	
(b) Differential conductance (d$I$/d$V$) calculated with and without including a lifetime broadening, $\Gamma(V)=\alpha |V|$, in the electrodes. The parameter $\alpha$ is fitted to the experimental results. }
	\label{conductance}
\end{figure}

The electronic transmission through a Pd adatom as a function of the energy
[Fig.~\ref{conductance}(a)] depends crucially on the tip-adatom distance $d$ but
for distances between $3$\,{\AA} and $2.6$\,{\AA} the typical conductances are on
the order of $(2-2.5)\,G_0$, which is in the range of the experimental values.
It is also of the same order of magnitude as found by Gava {\sl et al.} \cite{Gava2010} for the transmission between two  Pd$(001)$ surfaces connected by a small atomic chain.

The transmission through a magnetic Co adatom drops drastically (by a factor of two) compared to the case of the Pd adatom. For the sake of comparison  we have calculated the transmission through a hypothetical non-magnetic Co adatom for which the conductance at $E_{\mathrm F}$ is close to the one of the Pd adatom confirming the influence of the  local magnetization on the electronic transmission. These theoretical results are perfectly consistent with the experimental ones.

The complete modeling of the differential conductance curves is rather cumbersome since this would involve the calculation of the electrical current at various bias voltages and then calculating the derivative of $I(V)$. However in the present case, in the contact regime, for a highly symmetric system (identical leads) and in a very narrow voltage range around the Fermi level, one expects an almost odd $I(V)$ curve and therefore an even $dI/dV$ curve which can safely be approximated by the average transmission, $\frac{1}{2}[T(E\!=-\frac{V}{2})+T(E\!=\frac{V}{2})]$. Using this approximation we find flat and featureless $dI/dV$ spectra for bias voltages between $-0.1$ and $0.1$\,V at tip-adatom distances between 2.5 and 3{\AA} for both, Co and Pd contact, which do not reproduce the experimental findings.

\vspace{1cm}

The discrepancy between the experimentally observed and calculated conductance spectra can be lifted when taking the finite lifetime of the hot quasiparticles
into account. The quasiparticle lifetime in Pd has been calculated using various GW-based many-body perturbation theories\cite{Ladstadter2004,Zhukov2005} showing considerably shorter lifetimes close to $E_{F}$ compared to, e.\,g., Au or Cu. It is also displaying a different behavior compared to the free-electron behavior in Fermi liquid theory, $\hbar/\tau=\Gamma \propto (E-E_\mathrm F)^{2}$.
This is attributed to the role of the $d$-electrons and their finite bandwidth. Early calculations\cite{Longo1989} using a finite bandwidth model for an almost full band suggested an electronic level broadening due to spin-fluctuations in Pd of $\Gamma(E)\approx 0.05|E-E_\mathrm F|$. Lifetimes $\tau$ calculated within the GW approximation\cite{Zhukov2005} result in $\tau\sim 10$\,fs corresponding to $\Gamma\approx 50$\,meV for $E-E_\mathrm F\approx 250$meV.
Based on this we add an additional bias-dependent contribution to the imaginary part of the self-energy, $\Gamma(V)=\alpha |eV|$, to the $d$-orbitals in the two outermost surface layers in the calculation. This corresponds to the lifetime of electrons injected from the negative electrode that enter $eV$ above the Fermi level in the positive electrode. If we fit the unknown $\alpha$ to the experiments, as shown in Fig.~\ref{fig3}(d), we obtain $\alpha\sim 0.7$ corresponding to a lifetime of roughly 10\,fs at a bias of 100\,mV slightly below the GW results. 


It is interesting to note that similar spectroscopic features have been observed in mesoscopic contacts with conductances on the order of about 100\,$G_{0}$ and 800\,$G_{0}$, respectively \cite{Csonka2004, Ienaga2015}. While the details might differ, we expect that the mechanism leading to the suppression of conductance with increasing bias voltage in those contacts is the same.

Based on the theoretical calculations we attribute the decrease of the conductance with increasing bias to a suppression of the transmission of electrons. With increasing energy, more paramagnons are excited and more charge carriers are reflected. A characteristic feature of paramagnons is that they are overdamped modes, meaning they have short lifetimes and couple strongly to electrons to release their energy. As a consequence, their response in momentum space is broad and not distinctly defined giving rise to a broad and smeared out feature similar to what we observe. The strongest argument for the presence of paramagnons in Palladium is the extremely short lifetime of quasiparticles that is obtained from the fit to the observed spectra.
The different slopes of the d$I$/d$V$ spectra for Pd and Co contacts depicted in Fig.~4(b) are directly related to the magnetism of the Co adatom. While for palladium, the d-states at the Fermi level contribute substantially to the conductance, for cobalt the exchange coupling pushes the  majority-$d$-states below the Fermi level, and thus it is only spin-majority states of s-character and spin-minority states of d-character which contribute to the transport. The electronic states of $s$-character are only weakly affected by paramagnetic excitations, and therefore the $\Lambda$-shape is suppressed.

In conclusion, we have studied atomic contacts consisting of single Pd and Co atoms. Contacts of single Pd atoms exhibit pronounced $\Lambda$-shaped spetra, which we attribute to strong electron scattering near the contact which limits the lifetime of the charge carriers. This effect is explained by the presence of paramagnons. Conversely, contacts consisting of Co adatoms, which locally induce magnetic order, do not exhibit a similar suppression of conductance. Hence, we have demonstrated that signatures of paramagnons, which were commonly investigated by means of scattering techniques that integrate over larger areas, can be detected with local probes. This might expand the experimental tools for the investigation of high-temperatures superconductors.

\section{Acknowledgement}
VS, US, BB acknowledge funding by the SFB 767 and the Emmy-Noether-Program of the Deutsche Forschungsgemeinschaft. CB and MB would like to thank Derek Stewart for providing the pseudo-potential of palladium.

\bibliographystyle{apsrev4-1}

\begin{thebibliography}{39}%
\makeatletter
\providecommand \@ifxundefined [1]{%
 \@ifx{#1\undefined}
}%
\providecommand \@ifnum [1]{%
 \ifnum #1\expandafter \@firstoftwo
 \else \expandafter \@secondoftwo
 \fi
}%
\providecommand \@ifx [1]{%
 \ifx #1\expandafter \@firstoftwo
 \else \expandafter \@secondoftwo
 \fi
}%
\providecommand \natexlab [1]{#1}%
\providecommand \enquote  [1]{``#1''}%
\providecommand \bibnamefont  [1]{#1}%
\providecommand \bibfnamefont [1]{#1}%
\providecommand \citenamefont [1]{#1}%
\providecommand \href@noop [0]{\@secondoftwo}%
\providecommand \href [0]{\begingroup \@sanitize@url \@href}%
\providecommand \@href[1]{\@@startlink{#1}\@@href}%
\providecommand \@@href[1]{\endgroup#1\@@endlink}%
\providecommand \@sanitize@url [0]{\catcode `\\12\catcode `\$12\catcode
  `\&12\catcode `\#12\catcode `\^12\catcode `\_12\catcode `\%12\relax}%
\providecommand \@@startlink[1]{}%
\providecommand \@@endlink[0]{}%
\providecommand \url  [0]{\begingroup\@sanitize@url \@url }%
\providecommand \@url [1]{\endgroup\@href {#1}{\urlprefix }}%
\providecommand \urlprefix  [0]{URL }%
\providecommand \Eprint [0]{\href }%
\providecommand \doibase [0]{http://dx.doi.org/}%
\providecommand \selectlanguage [0]{\@gobble}%
\providecommand \bibinfo  [0]{\@secondoftwo}%
\providecommand \bibfield  [0]{\@secondoftwo}%
\providecommand \translation [1]{[#1]}%
\providecommand \BibitemOpen [0]{}%
\providecommand \bibitemStop [0]{}%
\providecommand \bibitemNoStop [0]{.\EOS\space}%
\providecommand \EOS [0]{\spacefactor3000\relax}%
\providecommand \BibitemShut  [1]{\csname bibitem#1\endcsname}%
\let\auto@bib@innerbib\@empty
\bibitem [{\citenamefont {Fay}\ and\ \citenamefont {Appel}(1980)}]{FayPRB}%
  \BibitemOpen
  \bibfield  {author} {\bibinfo {author} {\bibfnamefont {D.}~\bibnamefont
  {Fay}}\ and\ \bibinfo {author} {\bibfnamefont {J.}~\bibnamefont {Appel}},\
  }\href@noop {} {\bibfield  {journal} {\bibinfo  {journal} {Phys. Rev. B}\
  }\textbf {\bibinfo {volume} {22}},\ \bibinfo {pages} {3173} (\bibinfo {year}
  {1980})}\BibitemShut {NoStop}%
\bibitem [{\citenamefont {Moriya}\ \emph {et~al.}(1990)\citenamefont {Moriya},
  \citenamefont {Takahashi},\ and\ \citenamefont {Ueda}}]{Moriya}%
  \BibitemOpen
  \bibfield  {author} {\bibinfo {author} {\bibfnamefont {T.}~\bibnamefont
  {Moriya}}, \bibinfo {author} {\bibfnamefont {Y.}~\bibnamefont {Takahashi}}, \
  and\ \bibinfo {author} {\bibfnamefont {K.}~\bibnamefont {Ueda}},\ }\href@noop
  {} {\bibfield  {journal} {\bibinfo  {journal} {Journal of the Physical
  Society of Japan}\ }\textbf {\bibinfo {volume} {59}},\ \bibinfo {pages}
  {2905} (\bibinfo {year} {1990})}\BibitemShut {NoStop}%
\bibitem [{\citenamefont {Mathur}\ \emph {et~al.}(1998)\citenamefont {Mathur},
  \citenamefont {Grosche}, \citenamefont {Julian}, \citenamefont {Walker},
  \citenamefont {Freye}, \citenamefont {Haselwimmer},\ and\ \citenamefont
  {Lonzarich}}]{mathur_magnetically_1998}%
  \BibitemOpen
  \bibfield  {author} {\bibinfo {author} {\bibfnamefont {N.~D.}\ \bibnamefont
  {Mathur}}, \bibinfo {author} {\bibfnamefont {F.~M.}\ \bibnamefont {Grosche}},
  \bibinfo {author} {\bibfnamefont {S.~R.}\ \bibnamefont {Julian}}, \bibinfo
  {author} {\bibfnamefont {I.~R.}\ \bibnamefont {Walker}}, \bibinfo {author}
  {\bibfnamefont {D.~M.}\ \bibnamefont {Freye}}, \bibinfo {author}
  {\bibfnamefont {R.~K.~W.}\ \bibnamefont {Haselwimmer}}, \ and\ \bibinfo
  {author} {\bibfnamefont {G.~G.}\ \bibnamefont {Lonzarich}},\ }\href
  {http://www.nature.com/nature/journal/v394/n6688/abs/394039a0.html}
  {\bibfield  {journal} {\bibinfo  {journal} {Nature}\ }\textbf {\bibinfo
  {volume} {394}},\ \bibinfo {pages} {39} (\bibinfo {year} {1998})}\BibitemShut
  {NoStop}%
\bibitem [{\citenamefont {Scalapino}(2012)}]{scalapino_common_2012}%
  \BibitemOpen
  \bibfield  {author} {\bibinfo {author} {\bibfnamefont {D.~J.}\ \bibnamefont
  {Scalapino}},\ }\href {\doibase 10.1103/RevModPhys.84.1383} {\bibfield
  {journal} {\bibinfo  {journal} {Rev. Mod. Phys.}\ }\textbf {\bibinfo {volume}
  {84}},\ \bibinfo {pages} {1383} (\bibinfo {year} {2012})}\BibitemShut
  {NoStop}%
\bibitem [{\citenamefont {Stritzker}(1979)}]{Stritzker}%
  \BibitemOpen
  \bibfield  {author} {\bibinfo {author} {\bibfnamefont {B.}~\bibnamefont
  {Stritzker}},\ }\href@noop {} {\bibfield  {journal} {\bibinfo  {journal}
  {Phys. Rev. Lett.}\ }\textbf {\bibinfo {volume} {42}},\ \bibinfo {pages}
  {1769} (\bibinfo {year} {1979})}\BibitemShut {NoStop}%
\bibitem [{\citenamefont {Lonzarich}\ and\ \citenamefont
  {Taillefer}(1985)}]{Lonzarich1985}%
  \BibitemOpen
  \bibfield  {author} {\bibinfo {author} {\bibfnamefont {G.~G.}\ \bibnamefont
  {Lonzarich}}\ and\ \bibinfo {author} {\bibfnamefont {L.}~\bibnamefont
  {Taillefer}},\ }\href {\doibase 10.1088/0022-3719/18/22/017} {\bibfield
  {journal} {\bibinfo  {journal} {Journal of Physics C: Solid State Physics}\
  }\textbf {\bibinfo {volume} {18}},\ \bibinfo {pages} {4339} (\bibinfo {year}
  {1985})}\BibitemShut {NoStop}%
\bibitem [{\citenamefont {van Leeuwen}\ \emph {et~al.}(1992)\citenamefont {van
  Leeuwen}, \citenamefont {van Ruitenbeek}, \citenamefont {Schmid},\ and\
  \citenamefont {de~Jongh}}]{VanLeeuwen1992}%
  \BibitemOpen
  \bibfield  {author} {\bibinfo {author} {\bibfnamefont {D.~A.}\ \bibnamefont
  {van Leeuwen}}, \bibinfo {author} {\bibfnamefont {J.~M.}\ \bibnamefont {van
  Ruitenbeek}}, \bibinfo {author} {\bibfnamefont {G.}~\bibnamefont {Schmid}}, \
  and\ \bibinfo {author} {\bibfnamefont {L.}~\bibnamefont {de~Jongh}},\ }\href
  {\doibase 10.1016/0375-9601(92)90263-L} {\bibfield  {journal} {\bibinfo
  {journal} {Physics Letters A}\ }\textbf {\bibinfo {volume} {170}},\ \bibinfo
  {pages} {325} (\bibinfo {year} {1992})}\BibitemShut {NoStop}%
\bibitem [{\citenamefont {Mueller}\ \emph {et~al.}(1970)\citenamefont
  {Mueller}, \citenamefont {Freeman}, \citenamefont {Dimmock},\ and\
  \citenamefont {Furdyna}}]{Mueller1970}%
  \BibitemOpen
  \bibfield  {author} {\bibinfo {author} {\bibfnamefont {F.~M.}\ \bibnamefont
  {Mueller}}, \bibinfo {author} {\bibfnamefont {A.~J.}\ \bibnamefont
  {Freeman}}, \bibinfo {author} {\bibfnamefont {J.~O.}\ \bibnamefont
  {Dimmock}}, \ and\ \bibinfo {author} {\bibfnamefont {A.~M.}\ \bibnamefont
  {Furdyna}},\ }\href {\doibase 10.1103/PhysRevB.27.5116} {\bibfield  {journal}
  {\bibinfo  {journal} {Phys. Rev. B}\ }\textbf {\bibinfo {volume} {1}},\
  \bibinfo {pages} {4617} (\bibinfo {year} {1970})}\BibitemShut {NoStop}%
\bibitem [{\citenamefont {Doubble}\ \emph {et~al.}(2010)\citenamefont
  {Doubble}, \citenamefont {Hayden}, \citenamefont {Dai}, \citenamefont {Mook},
  \citenamefont {Thompson},\ and\ \citenamefont {Frost}}]{Doubble2010}%
  \BibitemOpen
  \bibfield  {author} {\bibinfo {author} {\bibfnamefont {R.}~\bibnamefont
  {Doubble}}, \bibinfo {author} {\bibfnamefont {S.~M.}\ \bibnamefont {Hayden}},
  \bibinfo {author} {\bibfnamefont {P.}~\bibnamefont {Dai}}, \bibinfo {author}
  {\bibfnamefont {H.~a.}\ \bibnamefont {Mook}}, \bibinfo {author}
  {\bibfnamefont {J.~R.}\ \bibnamefont {Thompson}}, \ and\ \bibinfo {author}
  {\bibfnamefont {C.~D.}\ \bibnamefont {Frost}},\ }\href {\doibase
  10.1103/PhysRevLett.105.027207} {\bibfield  {journal} {\bibinfo  {journal}
  {Phys. Rev. Lett.}\ }\textbf {\bibinfo {volume} {105}},\ \bibinfo {pages}
  {027207} (\bibinfo {year} {2010})}\BibitemShut {NoStop}%
\bibitem [{\citenamefont {Hayashi}\ \emph {et~al.}(2013)\citenamefont
  {Hayashi}, \citenamefont {Shimada}, \citenamefont {Jiang}, \citenamefont
  {Iwasawa}, \citenamefont {Aiura}, \citenamefont {Oguchi}, \citenamefont
  {Namatame},\ and\ \citenamefont {Taniguchi}}]{Hayashi2013}%
  \BibitemOpen
  \bibfield  {author} {\bibinfo {author} {\bibfnamefont {H.}~\bibnamefont
  {Hayashi}}, \bibinfo {author} {\bibfnamefont {K.}~\bibnamefont {Shimada}},
  \bibinfo {author} {\bibfnamefont {J.}~\bibnamefont {Jiang}}, \bibinfo
  {author} {\bibfnamefont {H.}~\bibnamefont {Iwasawa}}, \bibinfo {author}
  {\bibfnamefont {Y.}~\bibnamefont {Aiura}}, \bibinfo {author} {\bibfnamefont
  {T.}~\bibnamefont {Oguchi}}, \bibinfo {author} {\bibfnamefont
  {H.}~\bibnamefont {Namatame}}, \ and\ \bibinfo {author} {\bibfnamefont
  {M.}~\bibnamefont {Taniguchi}},\ }\href {\doibase 10.1103/PhysRevB.87.035140}
  {\bibfield  {journal} {\bibinfo  {journal} {Phys. Rev. B}\ }\textbf {\bibinfo
  {volume} {87}},\ \bibinfo {pages} {035140} (\bibinfo {year}
  {2013})}\BibitemShut {NoStop}%
\bibitem [{\citenamefont {Sushkov}(2014)}]{sushkov2014}%
  \BibitemOpen
  \bibfield  {author} {\bibinfo {author} {\bibfnamefont {O.~P.}\ \bibnamefont
  {Sushkov}},\ }\href {\doibase 10.2104/mbr07001} {\bibfield  {journal}
  {\bibinfo  {journal} {Nature Physics}\ }\textbf {\bibinfo {volume} {10}},\
  \bibinfo {pages} {339} (\bibinfo {year} {2014})}\BibitemShut {NoStop}%
\bibitem [{\citenamefont {Moriya}\ and\ \citenamefont
  {Takahashi}(1978)}]{Moriya1978}%
  \BibitemOpen
  \bibfield  {author} {\bibinfo {author} {\bibfnamefont {T.}~\bibnamefont
  {Moriya}}\ and\ \bibinfo {author} {\bibfnamefont {Y.}~\bibnamefont
  {Takahashi}},\ }\href {\doibase 10.1051/jphyscol:19786588} {\bibfield
  {journal} {\bibinfo  {journal} {Le Journal de Physique Colloques}\ }\textbf
  {\bibinfo {volume} {39}},\ \bibinfo {pages} {C6} (\bibinfo {year}
  {1978})}\BibitemShut {NoStop}%
\bibitem [{\citenamefont {Teng}\ \emph {et~al.}(2008)\citenamefont {Teng},
  \citenamefont {Han}, \citenamefont {Ku},\ and\ \citenamefont
  {H\"{u}cker}}]{Teng2008}%
  \BibitemOpen
  \bibfield  {author} {\bibinfo {author} {\bibfnamefont {X.}~\bibnamefont
  {Teng}}, \bibinfo {author} {\bibfnamefont {W.~Q.}\ \bibnamefont {Han}},
  \bibinfo {author} {\bibfnamefont {W.}~\bibnamefont {Ku}}, \ and\ \bibinfo
  {author} {\bibfnamefont {M.}~\bibnamefont {H\"{u}cker}},\ }\href {\doibase
  10.1002/anie.200704707} {\bibfield  {journal} {\bibinfo  {journal}
  {Angewandte Chemie - International Edition}\ }\textbf {\bibinfo {volume}
  {47}},\ \bibinfo {pages} {2055} (\bibinfo {year} {2008})}\BibitemShut
  {NoStop}%
\bibitem [{\citenamefont {Xiao}\ \emph {et~al.}(2009)\citenamefont {Xiao},
  \citenamefont {Ding}, \citenamefont {Shen}, \citenamefont {Yang},
  \citenamefont {Hui},\ and\ \citenamefont {Gao}}]{Xiao2009}%
  \BibitemOpen
  \bibfield  {author} {\bibinfo {author} {\bibfnamefont {C.}~\bibnamefont
  {Xiao}}, \bibinfo {author} {\bibfnamefont {H.}~\bibnamefont {Ding}}, \bibinfo
  {author} {\bibfnamefont {C.}~\bibnamefont {Shen}}, \bibinfo {author}
  {\bibfnamefont {T.}~\bibnamefont {Yang}}, \bibinfo {author} {\bibfnamefont
  {C.}~\bibnamefont {Hui}}, \ and\ \bibinfo {author} {\bibfnamefont {H.~J.}\
  \bibnamefont {Gao}},\ }\href {\doibase 10.1021/jp902005j} {\bibfield
  {journal} {\bibinfo  {journal} {Journal of Physical Chemistry C}\ }\textbf
  {\bibinfo {volume} {113}},\ \bibinfo {pages} {13466} (\bibinfo {year}
  {2009})}\BibitemShut {NoStop}%
\bibitem [{\citenamefont {Rodrigues}\ \emph {et~al.}(2003)\citenamefont
  {Rodrigues}, \citenamefont {Bettini}, \citenamefont {Silva},\ and\
  \citenamefont {Ugarte}}]{Rodrigues2003}%
  \BibitemOpen
  \bibfield  {author} {\bibinfo {author} {\bibfnamefont {V.}~\bibnamefont
  {Rodrigues}}, \bibinfo {author} {\bibfnamefont {J.}~\bibnamefont {Bettini}},
  \bibinfo {author} {\bibfnamefont {P.~C.}\ \bibnamefont {Silva}}, \ and\
  \bibinfo {author} {\bibfnamefont {D.}~\bibnamefont {Ugarte}},\ }\href
  {\doibase 10.1103/PhysRevLett.91.096801} {\bibfield  {journal} {\bibinfo
  {journal} {Phys. Rev. Lett.}\ }\textbf {\bibinfo {volume} {91}},\ \bibinfo
  {pages} {096801} (\bibinfo {year} {2003})},\ \Eprint
  {http://arxiv.org/abs/0307284} {arXiv:0307284 [cond-mat]} \BibitemShut
  {NoStop}%
\bibitem [{\citenamefont {Ienaga}\ \emph {et~al.}(2015)\citenamefont {Ienaga},
  \citenamefont {Takata}, \citenamefont {Onishi}, \citenamefont {Inagaki},
  \citenamefont {Tsujii}, \citenamefont {Kimura}, \citenamefont {Kawae},
  \citenamefont {Ienaga}, \citenamefont {Takata}, \citenamefont {Onishi},
  \citenamefont {Inagaki}, \citenamefont {Tsujii}, \citenamefont {Kimura},\
  and\ \citenamefont {Kawae}}]{Ienaga2015}%
  \BibitemOpen
  \bibfield  {author} {\bibinfo {author} {\bibfnamefont {K.}~\bibnamefont
  {Ienaga}}, \bibinfo {author} {\bibfnamefont {H.}~\bibnamefont {Takata}},
  \bibinfo {author} {\bibfnamefont {Y.}~\bibnamefont {Onishi}}, \bibinfo
  {author} {\bibfnamefont {Y.}~\bibnamefont {Inagaki}}, \bibinfo {author}
  {\bibfnamefont {H.}~\bibnamefont {Tsujii}}, \bibinfo {author} {\bibfnamefont
  {T.}~\bibnamefont {Kimura}}, \bibinfo {author} {\bibfnamefont
  {T.}~\bibnamefont {Kawae}}, \bibinfo {author} {\bibfnamefont
  {K.}~\bibnamefont {Ienaga}}, \bibinfo {author} {\bibfnamefont
  {H.}~\bibnamefont {Takata}}, \bibinfo {author} {\bibfnamefont
  {Y.}~\bibnamefont {Onishi}}, \bibinfo {author} {\bibfnamefont
  {Y.}~\bibnamefont {Inagaki}}, \bibinfo {author} {\bibfnamefont
  {H.}~\bibnamefont {Tsujii}}, \bibinfo {author} {\bibfnamefont
  {T.}~\bibnamefont {Kimura}}, \ and\ \bibinfo {author} {\bibfnamefont
  {T.}~\bibnamefont {Kawae}},\ }\href {\doibase 10.1063/1.4905729} {\bibfield
  {journal} {\bibinfo  {journal} {Applied Physics Letters}\ }\textbf {\bibinfo
  {volume} {106}},\ \bibinfo {pages} {2} (\bibinfo {year} {2015})}\BibitemShut
  {NoStop}%
\bibitem [{\citenamefont {Lee}(1998)}]{Lee1998}%
  \BibitemOpen
  \bibfield  {author} {\bibinfo {author} {\bibfnamefont {K.}~\bibnamefont
  {Lee}},\ }\href {\doibase 10.1103/PhysRevB.58.2391} {\bibfield  {journal}
  {\bibinfo  {journal} {Phys. Rev. B}\ }\textbf {\bibinfo {volume} {58}},\
  \bibinfo {pages} {2391} (\bibinfo {year} {1998})}\BibitemShut {NoStop}%
\bibitem [{\citenamefont {Delin}\ \emph {et~al.}(2003)\citenamefont {Delin},
  \citenamefont {Tosatti},\ and\ \citenamefont {Weht}}]{Delin2003}%
  \BibitemOpen
  \bibfield  {author} {\bibinfo {author} {\bibfnamefont {A.}~\bibnamefont
  {Delin}}, \bibinfo {author} {\bibfnamefont {E.}~\bibnamefont {Tosatti}}, \
  and\ \bibinfo {author} {\bibfnamefont {R.}~\bibnamefont {Weht}},\ }\href@noop
  {} {\bibfield  {journal} {\bibinfo  {journal} {Phys. Rev. Lett.}\ }\textbf
  {\bibinfo {volume} {92}},\ \bibinfo {pages} {057201} (\bibinfo {year}
  {2003})}\BibitemShut {NoStop}%
\bibitem [{\citenamefont {Smelova}\ \emph {et~al.}(2008)\citenamefont
  {Smelova}, \citenamefont {Bazhanov}, \citenamefont {Stepanyuk}, \citenamefont
  {Hergert}, \citenamefont {Saletsky},\ and\ \citenamefont
  {Bruno}}]{Smelova2008}%
  \BibitemOpen
  \bibfield  {author} {\bibinfo {author} {\bibfnamefont {K.~M.}\ \bibnamefont
  {Smelova}}, \bibinfo {author} {\bibfnamefont {D.~I.}\ \bibnamefont
  {Bazhanov}}, \bibinfo {author} {\bibfnamefont {V.~S.}\ \bibnamefont
  {Stepanyuk}}, \bibinfo {author} {\bibfnamefont {W.}~\bibnamefont {Hergert}},
  \bibinfo {author} {\bibfnamefont {A.~M.}\ \bibnamefont {Saletsky}}, \ and\
  \bibinfo {author} {\bibfnamefont {P.}~\bibnamefont {Bruno}},\ }\href
  {\doibase 10.1103/PhysRevB.77.033408} {\bibfield  {journal} {\bibinfo
  {journal} {Phys. Rev. B}\ }\textbf {\bibinfo {volume} {77}},\ \bibinfo
  {pages} {033408} (\bibinfo {year} {2008})}\BibitemShut {NoStop}%
\bibitem [{\citenamefont {Kudasov}\ and\ \citenamefont
  {Korshunov}(2007)}]{Kudasov2007}%
  \BibitemOpen
  \bibfield  {author} {\bibinfo {author} {\bibfnamefont {Y.~B.}\ \bibnamefont
  {Kudasov}}\ and\ \bibinfo {author} {\bibfnamefont {A.~S.}\ \bibnamefont
  {Korshunov}},\ }\href {\doibase 10.1016/j.physleta.2006.12.005} {\bibfield
  {journal} {\bibinfo  {journal} {Physics Letters, Section A: General, Atomic
  and Solid State Physics}\ }\textbf {\bibinfo {volume} {364}},\ \bibinfo
  {pages} {348} (\bibinfo {year} {2007})}\BibitemShut {NoStop}%
\bibitem [{\citenamefont {Sun}\ \emph {et~al.}(2010)\citenamefont {Sun},
  \citenamefont {Burton},\ and\ \citenamefont {Tsymbal}}]{Sun2010}%
  \BibitemOpen
  \bibfield  {author} {\bibinfo {author} {\bibfnamefont {Y.}~\bibnamefont
  {Sun}}, \bibinfo {author} {\bibfnamefont {J.~D.}\ \bibnamefont {Burton}}, \
  and\ \bibinfo {author} {\bibfnamefont {E.~Y.}\ \bibnamefont {Tsymbal}},\
  }\href {\doibase 10.1103/PhysRevB.81.064413} {\bibfield  {journal} {\bibinfo
  {journal} {Phys. Rev. B}\ }\textbf {\bibinfo {volume} {81}},\ \bibinfo
  {pages} {064413} (\bibinfo {year} {2010})}\BibitemShut {NoStop}%
\bibitem [{\citenamefont {Gava}\ \emph {et~al.}(2010)\citenamefont {Gava},
  \citenamefont {{Dal Corso}}, \citenamefont {Smogunov},\ and\ \citenamefont
  {Tosatti}}]{Gava2010}%
  \BibitemOpen
  \bibfield  {author} {\bibinfo {author} {\bibfnamefont {P.}~\bibnamefont
  {Gava}}, \bibinfo {author} {\bibfnamefont {A.}~\bibnamefont {{Dal Corso}}},
  \bibinfo {author} {\bibfnamefont {A.}~\bibnamefont {Smogunov}}, \ and\
  \bibinfo {author} {\bibfnamefont {E.}~\bibnamefont {Tosatti}},\ }\href
  {\doibase 10.1140/epjb/e2010-00046-1} {\bibfield  {journal} {\bibinfo
  {journal} {European Physical Journal B}\ }\textbf {\bibinfo {volume} {75}},\
  \bibinfo {pages} {57} (\bibinfo {year} {2010})}\BibitemShut {NoStop}%
\bibitem [{\citenamefont {Binnig}\ \emph {et~al.}(1982)\citenamefont {Binnig},
  \citenamefont {Rohrer}, \citenamefont {Gerber},\ and\ \citenamefont
  {Weibel}}]{Binnig1982}%
  \BibitemOpen
  \bibfield  {author} {\bibinfo {author} {\bibfnamefont {G.}~\bibnamefont
  {Binnig}}, \bibinfo {author} {\bibfnamefont {H.}~\bibnamefont {Rohrer}},
  \bibinfo {author} {\bibfnamefont {C.}~\bibnamefont {Gerber}}, \ and\ \bibinfo
  {author} {\bibfnamefont {E.}~\bibnamefont {Weibel}},\ }\href {\doibase
  10.1063/1.92999} {\bibfield  {journal} {\bibinfo  {journal} {Applied Physics
  Letters}\ }\textbf {\bibinfo {volume} {40}},\ \bibinfo {pages} {178}
  (\bibinfo {year} {1982})}\BibitemShut {NoStop}%
\bibitem [{\citenamefont {N\'eel}\ \emph {et~al.}(2007)\citenamefont {N\'eel},
  \citenamefont {Kr\"oger}, \citenamefont {Limot}, \citenamefont {Palotas},
  \citenamefont {Hofer},\ and\ \citenamefont {Berndt}}]{Neel}%
  \BibitemOpen
  \bibfield  {author} {\bibinfo {author} {\bibfnamefont {N.}~\bibnamefont
  {N\'eel}}, \bibinfo {author} {\bibfnamefont {J.}~\bibnamefont {Kr\"oger}},
  \bibinfo {author} {\bibfnamefont {L.}~\bibnamefont {Limot}}, \bibinfo
  {author} {\bibfnamefont {K.}~\bibnamefont {Palotas}}, \bibinfo {author}
  {\bibfnamefont {W.~A.}\ \bibnamefont {Hofer}}, \ and\ \bibinfo {author}
  {\bibfnamefont {R.}~\bibnamefont {Berndt}},\ }\href@noop {} {\bibfield
  {journal} {\bibinfo  {journal} {Phys. Rev. Lett.}\ }\textbf {\bibinfo
  {volume} {98}},\ \bibinfo {pages} {016801} (\bibinfo {year}
  {2007})}\BibitemShut {NoStop}%
\bibitem [{\citenamefont {Vitali}\ \emph {et~al.}(2008)\citenamefont {Vitali},
  \citenamefont {Ohmann}, \citenamefont {Stepanow}, \citenamefont
  {Gambardella}, \citenamefont {Tao}, \citenamefont {Huang}, \citenamefont
  {Stepanyuk}, \citenamefont {Bruno},\ and\ \citenamefont {Kern}}]{Vitali}%
  \BibitemOpen
  \bibfield  {author} {\bibinfo {author} {\bibfnamefont {L.}~\bibnamefont
  {Vitali}}, \bibinfo {author} {\bibfnamefont {R.}~\bibnamefont {Ohmann}},
  \bibinfo {author} {\bibfnamefont {S.}~\bibnamefont {Stepanow}}, \bibinfo
  {author} {\bibfnamefont {P.}~\bibnamefont {Gambardella}}, \bibinfo {author}
  {\bibfnamefont {K.}~\bibnamefont {Tao}}, \bibinfo {author} {\bibfnamefont
  {R.}~\bibnamefont {Huang}}, \bibinfo {author} {\bibfnamefont {V.~S.}\
  \bibnamefont {Stepanyuk}}, \bibinfo {author} {\bibfnamefont {P.}~\bibnamefont
  {Bruno}}, \ and\ \bibinfo {author} {\bibfnamefont {K.}~\bibnamefont {Kern}},\
  }\href@noop {} {\bibfield  {journal} {\bibinfo  {journal} {Phys. Rev. Lett.}\
  }\textbf {\bibinfo {volume} {101}},\ \bibinfo {pages} {216802} (\bibinfo
  {year} {2008})}\BibitemShut {NoStop}%
\bibitem [{\citenamefont {Csonka}\ \emph {et~al.}(2004)\citenamefont {Csonka},
  \citenamefont {Halbritter}, \citenamefont {Mih\'{a}ly}, \citenamefont
  {Shklyarevskii}, \citenamefont {Speller},\ and\ \citenamefont {{Van
  Kempen}}}]{Csonka2004}%
  \BibitemOpen
  \bibfield  {author} {\bibinfo {author} {\bibfnamefont {S.}~\bibnamefont
  {Csonka}}, \bibinfo {author} {\bibfnamefont {A.}~\bibnamefont {Halbritter}},
  \bibinfo {author} {\bibfnamefont {G.}~\bibnamefont {Mih\'{a}ly}}, \bibinfo
  {author} {\bibfnamefont {O.~I.}\ \bibnamefont {Shklyarevskii}}, \bibinfo
  {author} {\bibfnamefont {S.}~\bibnamefont {Speller}}, \ and\ \bibinfo
  {author} {\bibfnamefont {H.}~\bibnamefont {{Van Kempen}}},\ }\href@noop {}
  {\bibfield  {journal} {\bibinfo  {journal} {Phys. Rev. Lett.}\ }\textbf
  {\bibinfo {volume} {93}},\ \bibinfo {pages} {016802} (\bibinfo {year}
  {2004})}\BibitemShut {NoStop}%
\bibitem [{\citenamefont {Matsuda}\ and\ \citenamefont
  {Kizuka}(2007)}]{Matsuda2007}%
  \BibitemOpen
  \bibfield  {author} {\bibinfo {author} {\bibfnamefont {T.}~\bibnamefont
  {Matsuda}}\ and\ \bibinfo {author} {\bibfnamefont {T.}~\bibnamefont
  {Kizuka}},\ }\href {\doibase 10.1143/JJAP.46.4370} {\bibfield  {journal}
  {\bibinfo  {journal} {Japanese Journal of Applied Physics,}\ }\textbf
  {\bibinfo {volume} {46}},\ \bibinfo {pages} {4370} (\bibinfo {year}
  {2007})}\BibitemShut {NoStop}%
\bibitem [{\citenamefont {Ternes}\ \emph {et~al.}(2009)\citenamefont {Ternes},
  \citenamefont {Heinrich},\ and\ \citenamefont {Schneider}}]{Ternes2009a}%
  \BibitemOpen
  \bibfield  {author} {\bibinfo {author} {\bibfnamefont {M.}~\bibnamefont
  {Ternes}}, \bibinfo {author} {\bibfnamefont {A.~J.}\ \bibnamefont
  {Heinrich}}, \ and\ \bibinfo {author} {\bibfnamefont {W.-D.}\ \bibnamefont
  {Schneider}},\ }\href {\doibase 10.1088/0953-8984/21/5/053001} {\bibfield
  {journal} {\bibinfo  {journal} {J. Phys.: Condens. Matter}\ }\textbf
  {\bibinfo {volume} {21}},\ \bibinfo {pages} {053001} (\bibinfo {year}
  {2009})}\BibitemShut {NoStop}%
\bibitem [{\citenamefont {Schneider}\ \emph {et~al.}(2005)\citenamefont
  {Schneider}, \citenamefont {Vitali}, \citenamefont {Wahl}, \citenamefont
  {Knorr}, \citenamefont {Diekhoner}, \citenamefont {Wittich}, \citenamefont
  {Vogelgesang},\ and\ \citenamefont {Kern}}]{Wahl2005}%
  \BibitemOpen
  \bibfield  {author} {\bibinfo {author} {\bibfnamefont {M.}~\bibnamefont
  {Schneider}}, \bibinfo {author} {\bibfnamefont {L.}~\bibnamefont {Vitali}},
  \bibinfo {author} {\bibfnamefont {P.}~\bibnamefont {Wahl}}, \bibinfo {author}
  {\bibfnamefont {N.}~\bibnamefont {Knorr}}, \bibinfo {author} {\bibfnamefont
  {L.}~\bibnamefont {Diekhoner}}, \bibinfo {author} {\bibfnamefont
  {G.}~\bibnamefont {Wittich}}, \bibinfo {author} {\bibfnamefont
  {M.}~\bibnamefont {Vogelgesang}}, \ and\ \bibinfo {author} {\bibfnamefont
  {K.}~\bibnamefont {Kern}},\ }\href {\doibase 10.1007/s00339-004-3119-7}
  {\bibfield  {journal} {\bibinfo  {journal} {Applied Physics A}\ }\textbf
  {\bibinfo {volume} {80}},\ \bibinfo {pages} {937} (\bibinfo {year}
  {2005})}\BibitemShut {NoStop}%
\bibitem [{qua()}]{quantumwise}%
  \BibitemOpen
  \href@noop {} {}\bibinfo {note} {Atomistix ToolKit version 2015.1,
  QuantumWise A/S (www.quantumwise.com)}\BibitemShut {NoStop}%
\bibitem [{\citenamefont {Brandbyge}\ \emph {et~al.}(2002)\citenamefont
  {Brandbyge}, \citenamefont {Mozos}, \citenamefont {Ordej\'{o}n},
  \citenamefont {Taylor},\ and\ \citenamefont {Stokbro}}]{Brandbyge2002}%
  \BibitemOpen
  \bibfield  {author} {\bibinfo {author} {\bibfnamefont {M.}~\bibnamefont
  {Brandbyge}}, \bibinfo {author} {\bibfnamefont {J.-L.}\ \bibnamefont
  {Mozos}}, \bibinfo {author} {\bibfnamefont {P.}~\bibnamefont {Ordej\'{o}n}},
  \bibinfo {author} {\bibfnamefont {J.}~\bibnamefont {Taylor}}, \ and\ \bibinfo
  {author} {\bibfnamefont {K.}~\bibnamefont {Stokbro}},\ }\href {\doibase
  10.1103/PhysRevB.65.165401} {\bibfield  {journal} {\bibinfo  {journal} {Phys.
  Rev. B}\ }\textbf {\bibinfo {volume} {65}},\ \bibinfo {pages} {165401}
  (\bibinfo {year} {2002})}\BibitemShut {NoStop}%
\bibitem [{\citenamefont {Perdew}(1981)}]{Perdew1981}%
  \BibitemOpen
  \bibfield  {author} {\bibinfo {author} {\bibfnamefont {J.~P.}\ \bibnamefont
  {Perdew}},\ }\href {\doibase 10.1103/PhysRevB.23.5048} {\bibfield  {journal}
  {\bibinfo  {journal} {Phys. Rev. B}\ }\textbf {\bibinfo {volume} {23}},\
  \bibinfo {pages} {5048} (\bibinfo {year} {1981})}\BibitemShut {NoStop}%
\bibitem [{\citenamefont {Alexandre}\ \emph {et~al.}(2006)\citenamefont
  {Alexandre}, \citenamefont {Mattesini}, \citenamefont {Soler},\ and\
  \citenamefont {Yndurain}}]{Alexandre2006b}%
  \BibitemOpen
  \bibfield  {author} {\bibinfo {author} {\bibfnamefont {S.~S.}\ \bibnamefont
  {Alexandre}}, \bibinfo {author} {\bibfnamefont {M.}~\bibnamefont
  {Mattesini}}, \bibinfo {author} {\bibfnamefont {J.~M.}\ \bibnamefont
  {Soler}}, \ and\ \bibinfo {author} {\bibfnamefont {F.}~\bibnamefont
  {Yndurain}},\ }\href {\doibase 10.1103/PhysRevLett.96.079701} {\bibfield
  {journal} {\bibinfo  {journal} {Phys. Rev. Lett.}\ }\textbf {\bibinfo
  {volume} {96}},\ \bibinfo {pages} {079701; author reply 079702} (\bibinfo
  {year} {2006})}\BibitemShut {NoStop}%
\bibitem [{Note1()}]{Note1}%
  \BibitemOpen
  \bibinfo {note} {Our LDA calculations predict an equilibrium lattice
  parameter of $3.87$\protect \tmspace +\thinmuskip {.1667em}{\r A} for face
  centered (fcc) Pd and an on-set of magnetism for lattice parameters above
  $3.96$\protect \tmspace +\thinmuskip {.1667em}{\r A} while GGA calculations
  predict a magnetic bulk at equilibrium.}\BibitemShut {Stop}%
\bibitem [{\citenamefont {Troullier}\ and\ \citenamefont
  {Martins}(1991)}]{Troullier1991}%
  \BibitemOpen
  \bibfield  {author} {\bibinfo {author} {\bibfnamefont {N.}~\bibnamefont
  {Troullier}}\ and\ \bibinfo {author} {\bibfnamefont {J.~L.}\ \bibnamefont
  {Martins}},\ }\href {\doibase 10.1103/PhysRevB.43.1993} {\bibfield  {journal}
  {\bibinfo  {journal} {Phys. Rev. B}\ }\textbf {\bibinfo {volume} {43}},\
  \bibinfo {pages} {1993} (\bibinfo {year} {1991})}\BibitemShut {NoStop}%
\bibitem [{Note2()}]{Note2}%
  \BibitemOpen
  \bibinfo {note} {Note that when considering the effect of a tip in contact
  with the adatom we have found that a rather pronounced dip develops around
  $E_\protect \mathrm F$ which will play strongly against the magnetization of
  the adatom.}\BibitemShut {Stop}%
\bibitem [{\citenamefont {Ladst{\"{a}}dter}\ \emph {et~al.}(2004)\citenamefont
  {Ladst{\"{a}}dter}, \citenamefont {Hohenester}, \citenamefont {Puschnig},\
  and\ \citenamefont {Ambrosch-Draxl}}]{Ladstadter2004}%
  \BibitemOpen
  \bibfield  {author} {\bibinfo {author} {\bibfnamefont {F.}~\bibnamefont
  {Ladst{\"{a}}dter}}, \bibinfo {author} {\bibfnamefont {U.}~\bibnamefont
  {Hohenester}}, \bibinfo {author} {\bibfnamefont {P.}~\bibnamefont
  {Puschnig}}, \ and\ \bibinfo {author} {\bibfnamefont {C.}~\bibnamefont
  {Ambrosch-Draxl}},\ }\href {\doibase 10.1103/PhysRevB.70.235125} {\bibfield
  {journal} {\bibinfo  {journal} {Phys. Rev. B}\ }\textbf {\bibinfo {volume}
  {70}},\ \bibinfo {pages} {235125} (\bibinfo {year} {2004})}\BibitemShut
  {NoStop}%
\bibitem [{\citenamefont {Zhukov}\ \emph {et~al.}(2005)\citenamefont {Zhukov},
  \citenamefont {Chulkov},\ and\ \citenamefont {Echenique}}]{Zhukov2005}%
  \BibitemOpen
  \bibfield  {author} {\bibinfo {author} {\bibfnamefont {V.~P.}\ \bibnamefont
  {Zhukov}}, \bibinfo {author} {\bibfnamefont {E.~V.}\ \bibnamefont {Chulkov}},
  \ and\ \bibinfo {author} {\bibfnamefont {P.~M.}\ \bibnamefont {Echenique}},\
  }\href {\doibase 10.1103/PhysRevB.72.155109} {\bibfield  {journal} {\bibinfo
  {journal} {Phys. Rev. B}\ }\textbf {\bibinfo {volume} {72}},\ \bibinfo
  {pages} {155109} (\bibinfo {year} {2005})}\BibitemShut {NoStop}%
\bibitem [{\citenamefont {Longo}\ and\ \citenamefont
  {Mitrovi{\'{c}}}(1989)}]{Longo1989}%
  \BibitemOpen
  \bibfield  {author} {\bibinfo {author} {\bibfnamefont {J.~P.}\ \bibnamefont
  {Longo}}\ and\ \bibinfo {author} {\bibfnamefont {B.}~\bibnamefont
  {Mitrovi{\'{c}}}},\ }\href {\doibase 10.1007/BF00681757} {\bibfield
  {journal} {\bibinfo  {journal} {J. Low Temp. Phys.}\ }\textbf {\bibinfo
  {volume} {74}},\ \bibinfo {pages} {141} (\bibinfo {year} {1989})}\BibitemShut
  {NoStop}%
\end{thebibliography}
%

\end{document}